\documentclass{tMOP2e}
\usepackage{amsmath,psfrag,graphics}
\newcommand{\Ns}{{N}_{\rm th}}

\begin{document}
\title{Constrained MaxLik reconstruction of multimode photon distributions}
\author{ G. Brida$^1$, M. Genovese$^1$, A. Meda $^1$, S. Olivares$^{2,3}$,
M. G. A. Paris$^{2,3,4}$, F. Piacentini$^1$\\
\vspace{3pt} $^{1}${\em{Istituto Nazionale di Ricerca Metrologica, I-10135 Torino, Italia}}\\
\vspace{3pt} $^{2}${\em{CNISM UdR Milano Universit\`a, I-20133 Milano, Italia}}\\
\vspace{3pt} $^{3}${\em{Dipartimento di Fisica, Universit\`a di Milano, I-20133 Milano, Italia}}\\
\vspace{3pt} $^{4}${\em{ISI Foundation, I-10133 Torino, Italia}}\\
\vspace{3pt}\received{xx/xx/xxxx}}
\maketitle
\begin{abstract}
We address the reconstruction of the full photon distribution of
multimode fields generated by seeded parametric down-conversion (PDC). Our
scheme is based on on/off avalanche photodetection assisted by
maximum-likelihood (MaxLik) estimation and does not involve photon
counting. We present a novel constrained MaxLik method that
incorporates the request of finite energy to improve the rate of
convergence and, in turn, the overall accuracy of the
reconstruction.
\end{abstract}
\section{Introduction}
\label{s:intro}
The reconstruction of photon statistics of quantum optical fields
is of the utmost relevance for several applications, ranging from
quantum information \cite{zei} to the foundations of quantum
mechanics \cite{MG} and quantum optics \cite{man1,man2,man3,man4,Mandel}.
Despite this fact, the realization of photodetectors well suited for
this purpose still represents an experimental challenge. The few
existing examples \cite{riv1,riv2,riv3,riv4,riv5,riv6} show severe
limitations. On the other hand, reconstruction schemes based on
quantum tomography \cite{mun,zhang,raymerLNP} require phase-matching
with a suitable local oscillator and do not represent a technique suited
for a diffuse use.
This situation prompted various theoretical studies \cite{mogy1,mogy2,
pcount,ar} addressed to achieve the reconstruction of the (diagonal)
elements of the density matrix exploiting the information achievable
with realistic detectors.
\par
In a recent series of papers \cite{nos1,nos2,nos3,nos4,nos5},  we
have demonstrated how a very satisfactory reconstruction of the
statistics of mono-partite and bi-partite quantum optical states may
be obtained using the simplest kind of detectors, namely on/off
detectors \cite{pcount,ar} operating in the Geiger mode, whose
outcomes are either ``off'' (no photons detected) or ``on'', i.e., a
``click'', indicating the detection of one or more photons. Our method
recovers the full photon statistics using maximum likelihood
(MaxLik) reconstruction on on/off data obtained using variable
detection efficiency (by inserting calibrated neutral filters).
\par
In this paper we present a modified version of our method, which
improves both convergence and accuracy upon incorporating the
obvious {\em a priori} constraint of finite signal energy. In
particular, we address the reconstruction of the full photon
distribution of multimode fields generated by seeded parametric
down-conversion. The novel method allows to overcome the increased
complexity of the reconstruction problem and represents an important
step in view of a widespread application of MaxLik reconstruction.
\par
The paper is structured as follows: in the next Section we describe the
constrained MaxLik algorithm in some details, whereas in Section
\ref{s:test} we describe the experimental apparatus and illustrate the
application of the method to the reconstruction of the photon
distribution of multimode fields generated by seeded parametric
down-conversion. Finally, Section \ref{s:remarks} closes the paper
with some concluding remarks.
\section{Constrained MaxLik algorithm}
\label{s:algorithm}
The probability $p_0(\eta)$ that a photodetector with quantum
efficiency $\eta$ does not click, when an input quantum state
$\varrho = \sum_{n,m} \varrho_{nm} |n\rangle\langle m|$
impinges on it, reads as follows:
\begin{equation}\label{p:off}
p_0(\eta) = \sum_{n} (1-\eta)^n \varrho_{n},
\end{equation}
where $\varrho_n = \varrho_{nn}$ is the $n$-th entry of the photon
distribution of the input state.
Now, if we consider a set of $N$ detectors with different quantum
efficiencies $\eta_\nu$, $\nu=1,\ldots,N$, then we can write the ``off''
probabilities as:
\begin{equation}\label{off:prob}
{\sf P}_\nu \equiv p_0(\eta_\nu) = \sum_{n} A_{\nu n} \varrho_{n},
\end{equation}
where $A_{\nu n} = (1-\eta_\nu)^n$. Looking at Eq.~(\ref{off:prob}) as
a statistical model for the parameters $\varrho_{n}$, we can solve
it by the MaxLik estimation. We proceed as follows: first
of all, we assume there exists a value $\tilde{n}$ such
that $\varrho_n$ is negligible for $n > \tilde{n}$;
we assign the {\em loglikelihood} function (with normalized ${\sf P}_\nu$),
that is the global probability of the sample:
\begin{equation}\label{LL:stand}
{\sf L} = \frac{1}{N_x}
\log \prod_\nu\left(
\frac{{\sf P}_\nu}{\sum_\lambda {\sf P}_\lambda}
\right)^{{\cal N}_\nu} = \sum_\nu f_\nu
\log\frac{{\sf P}_\nu}{\sum_\lambda {\sf P}_\lambda},
\end{equation}
where $f_\nu = {\cal N}_\nu / N_x$ is the experimental frequency of
``off'' events, ${\cal N}_\nu$ being the number of ``off'' events
for a fixed quantum efficiency $\eta_\nu$ and $N_x$ the total events
amount. The MaxLik estimated $\varrho_n$ values are the ones
maximizing ${\sf L}$. Since the model is linear and the unknowns
$\varrho_n$ are positive, we can find the expectation-maximization
solution of the MaxLik problem by means of an iterative procedure as
described in \cite{kon,mogy1,nos1,Hra,JCML}, i.e.:
\begin{equation}\label{EM:sol:st}
\varrho_n^{(h+1)} = \frac{\varrho_n^{(h)}}{\sum_m {\varrho_m^{(h)}}}
\sum_\nu \frac{A_{\nu n}}{(\sum_\lambda A_{\lambda n})}
\frac{f_\nu}{{\sf P}_{\nu}^{(h)}},
\end{equation}
where $\varrho_n^{(h)}$ is the value of $\varrho_n$ evaluated at the $h$-th
iteration, and ${\sf P}_\nu^{(h)} = \sum_n A_{\nu n}\,\varrho_n^{(h)}$.
The algorithm (\ref{EM:sol:st}), known to converge unbiasedly to the MaxLik
solution, provides a solution once the initial distribution $\varrho_n^{(0)}$
is chosen. On the other hand, the initial distribution slightly
affects only the convergence rate and \emph{not} the precision at
convergence \cite{pcount}.
\par
As a matter of fact, the solution obtained above corresponds to the
best photon distribution fitting the experimental data, i.e., the
measured ``off'' probabilities. However, it is possible that
different photon distributions fit the same experimental data,
giving rise to a {\em family} of {\em suitable} distributions. In
these cases it could happen that the MaxLik solution, even
if in good agreement with the available experimental knowledge, may be
different from the actual (unknown) one. Indeed, this is the case of
multimode fields when the number of modes grows. In order to overcome
this limitation, we developed a modified version of the MaxLik algorithm,
which incorporated the constraint of finite energy for the incoming
signal, i.e., the quantity $\sum_n n \varrho_n$. In practice,
we maximize the loglikelihood (\ref{LL:stand}) with a constraint
on the energy. Now the function to be maximized with respect to $\varrho_n$
is:
\begin{equation}
{\sf L}_\beta = {\sf L} - \beta \sum_n n \varrho_n,
\end{equation}
${\sf L}$ being given in (\ref{LL:stand}) and $\beta$ being a Lagrange
multiplier. The equations
$\frac{\partial {\sf L}_\beta}{\partial \varrho_n}=0$ lead to:
\begin{equation}\label{quasi}
\left(
\frac{\sum_\gamma {\sf P}_\gamma}{\sum_\mu f_\mu}
\right) \sum_\nu
\frac{A_{\nu n}}
{\left[\sum_\lambda A_{\lambda n} + \beta\, n \left(
\frac{\sum_{\gamma'} {\sf P}_{\gamma'}}{\sum_{\mu'} f_{\mu'}}
\right)\right]}
\frac{f_\nu}{{\sf P}_{\nu}} =1,
\end{equation}
and, then, by multiplying both the sides of Eq.~(\ref{quasi}) by
$\varrho_n$, we get a map ${\boldsymbol T} \varrho_n = \varrho_n$,
whose fixed point can be obtained by the following iterative solution:
\begin{equation}\label{EM:sol:WC}
\varrho_n^{(h+1)} =
\frac{\varrho_n^{(h)}}{\sum_m {\varrho_m^{(h)}}}
\sum_\nu \frac{A_{\nu n}}
{\left[\sum_\lambda A_{\lambda n} + \beta\, n \left(
\frac{\sum_{\gamma'} {\sf P}^{(h)}_{\gamma'}}{\sum_{\mu'} f_{\mu'}}
\right)\right]}
\frac{f_\nu}{{\sf P}^{(h)}_{\nu}}.
\end{equation}
The parameter $\beta$ can be tuned in order to control the energy
of the reconstructed state and improve both the convergence rate and
the overall accuracy. Of course, if we take $\beta=0$, then
Eq.s~(\ref{EM:sol:WC}) and (\ref{EM:sol:st}) become the same.
\par
Indeed, in order to use Eq.~(\ref{EM:sol:WC}) we need to know the
input state energy which, in general, cannot be directly accessible
from experimental data. However, in cases when a model of the photon
distribution of the input state is available, we can estimate
{\em indirectly} this energy by a simple fit of the ``off'' probabilities.
\par
In the following, we consider the multimode field obtained by seeded
Parametric Down Conversion (PDC). In this case, the photon
distribution is expected to have the form:
\begin{equation}\label{DTpd}
\varrho_{n} = \frac{(\Ns)^n}{(1+\Ns)^{n+M}}\,
\exp\left( -\frac{|\alpha|^2}{1+\Ns}\right)\,
L_n^{M-1}\left( -\frac{|\alpha|^2}{\Ns (1+\Ns)} \right),
\end{equation}
where $\varrho_{n}\equiv \varrho_{n}(\Ns,\alpha,M)$ and $L_n^a(z)$
are the generalized Laguerre polynomials.
Eq.~(\ref{DTpd}) represents the convolution of $M$ thermal
states ($M$ {\em spatial} modes), all with the same average number of
thermal photons $\Ns>0$ except for one, displaced by an
amount $\alpha$. This model will be justified by the experimental
setup described in Section \ref{s:test}. From Eq.s~(\ref{p:off}) and
(\ref{DTpd}) we can calculate the ``off'' probability
$p_0\equiv p_0(\Ns,\alpha,M,\eta)$, that is:
\begin{align}
p_0= \sum_{n=0}^{\infty} (1-\eta)^n\, \varrho_n(\Ns,\alpha,M)
= \frac{1}{(1+\eta \Ns)^M}\,
\exp \left(  -\frac{\eta |\alpha|^2}{1+\eta\Ns} \right),\label{P0convThD}
\end{align}
$\eta$ being the quantum efficiency of the on/off photodetector.
Notice that, from Eq.~(\ref{P0convThD}), one can obtain the relevant cases
for a Poissonian input photon distribution ($\Ns \to 0$) and for
a multithermal one ($\alpha \to 0$). Thanks to Eq.~(\ref{P0convThD}), one
can evaluate the input state energy $\Ns+|\alpha|^2$, and choose a
suitable value for $\beta$.
\par
Before the end of this section, it is worth pointing out that
if each spatial mode consists of $M'$ {\em temporal}
modes, then the input photon distribution and the ``off'' probability
are still given by Eq.s~(\ref{DTpd}) and (\ref{P0convThD}),
respectively, but with ${\cal M} \rightarrow M\times M'$ in place of $M$.
\section{Experimental test}\label{s:test}
\begin{figure}[htbp]
\begin{center}
\includegraphics[width=0.8\textwidth]{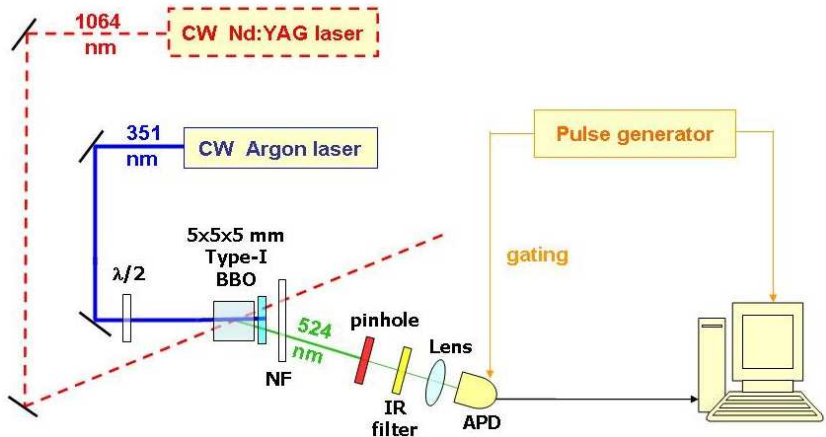}
\end{center}
\caption{Experimental setup: the stimulated emission at
$\lambda_{\rm stimul}=524$ nm is addressed to the NF and then collected by
the APD. The number and temporal width of the acquisition windows is set
by the pulse generator used for the detector's gating.}
\label{f:setup_stimul}
\end{figure}

In order to test the reliability of the algorithm reported in Eq.
(\ref{EM:sol:WC}), we applied it to the reconstruction of a
stimulated type-I PDC branch, at different stimulation regimes. In
our experiment, whose setup is shown in Fig.~\ref{f:setup_stimul}, a
CW Argon laser ($\lambda_{\rm pump}=351.1$ nm) pumps a
$5\times5\times5$ mm type-I BBO crystal, generating PDC. Together
with the pump beam, a CW Nd:Yag laser ($\lambda_{\rm seed}=1064$ nm)
is injected into the crystal in the proper way to generate
stimulated PDC, and we look at the emission in the ${\mathbf k}_{\rm
stimul}$ direction ($\lambda_{\rm stimul}=524$ nm).
Different values of quantum efficiency have been obtained by
inserting properly calibrated Schott neutral filters (NF), starting
from $\eta_{\rm max}=28.4\%$; after them, on the optical path of the
stimulated branch we put an anti-infrared (IR) filter (to cut off
the noise due to the Nd:Yag laser dispersion), a variable pinhole
(to control the number $M$ of spatial propagation modes collected)
and a fiber coupler connected by a multimode fiber with the detector
(avalanche photodiode, Perkin Elmer SPCM-AQR-15). We set the pulse
generator in order to open in the APD $2\,10^{5}$ detection windows
per second, each one of 20 ns; the pinhole diameter is regulated in
order to collect only few spatial modes (more precisely $M=7$), of
whom only one stimulated. Moreover, each spatial mode consists of
many temporal modes: the total number of modes can then be estimated
${\cal M}=7\,10^{5}$. It is worth to mention that, when the number of
modes exceeds few tens, the dependence on this parameter is rather
small and a rough estimate of the order of magnitude suffices.
\par
We have performed three separate data collections, each one corresponding to a
different stimulation regime: by indicating with $x$ the percentage of
stimulated emission on the whole PDC amount collected, our acquisitions
were respectively characterized by $x=51.4\%$, $x=78.1\%$ and $x=90.7\%$.
The evaluation of the background photons have been performed through an
acquisition step without PDC emission (Argon pump off, Nd:Yag seed on),
followed by a proper subtraction from data.
\begin{figure}[htbp]
\psfrag{P0}{\small $f_0$}
\psfrag{eta}{\small $\eta$}
\psfrag{pn}{\small $\varrho_n$}
\psfrag{n}{\small $n$}
\psfrag{Nave51}{\tiny $N_{\rm ave}=7.23$}
\psfrag{x51}{\tiny $x=50.7\%$}
\psfrag{chi51}{\tiny $\chi^2=4.1\, 10^{-2}$}
\psfrag{Fid51}{\tiny $F=99.57\%$}
\psfrag{Nave78}{\tiny $N_{\rm ave}=16.72$}
\psfrag{x78}{\tiny $x=78.1\%$}
\psfrag{chi78}{\tiny $\chi^2=3.4\, 10^{-2}$}
\psfrag{Fid78}{\tiny $F=99.97\%$}
\psfrag{Nave91}{\tiny $N_{\rm ave}=18.34$}
\psfrag{x91}{\tiny $x=90.7\%$}
\psfrag{chi91}{\tiny $\chi^2=9.1\, 10^{-3}$}
\psfrag{Fid91}{\tiny $F=99.98\%$}
\begin{center}
\includegraphics[width=0.45\textwidth]{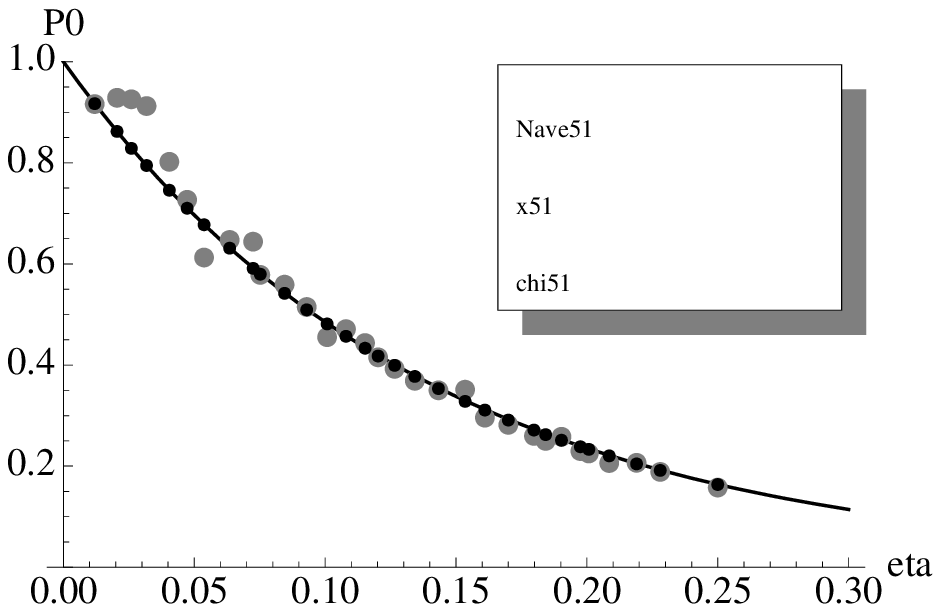}\hspace{0.5cm}
\includegraphics[width=0.45\textwidth]{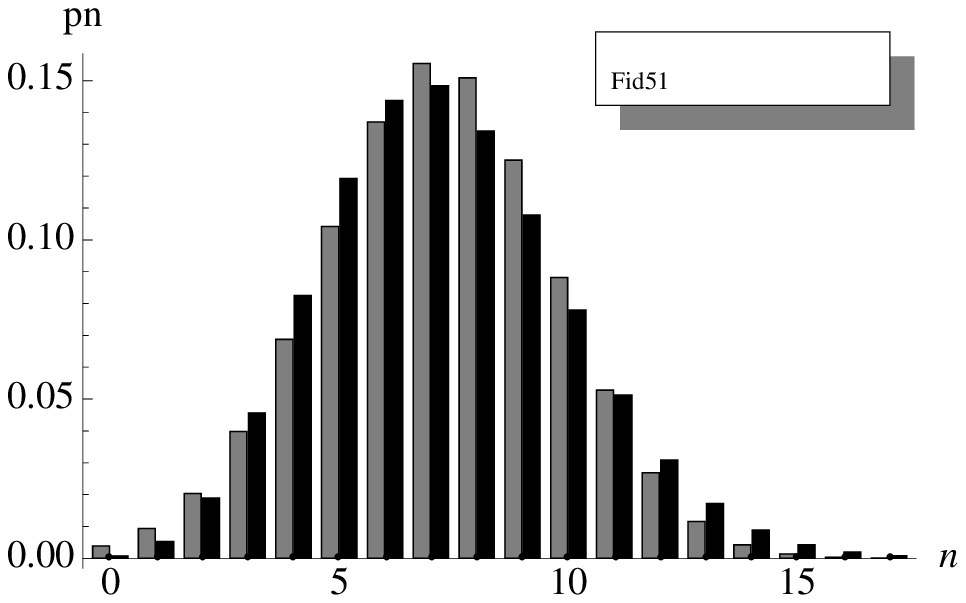}
\vspace{0.3cm}
\includegraphics[width=0.45\textwidth]{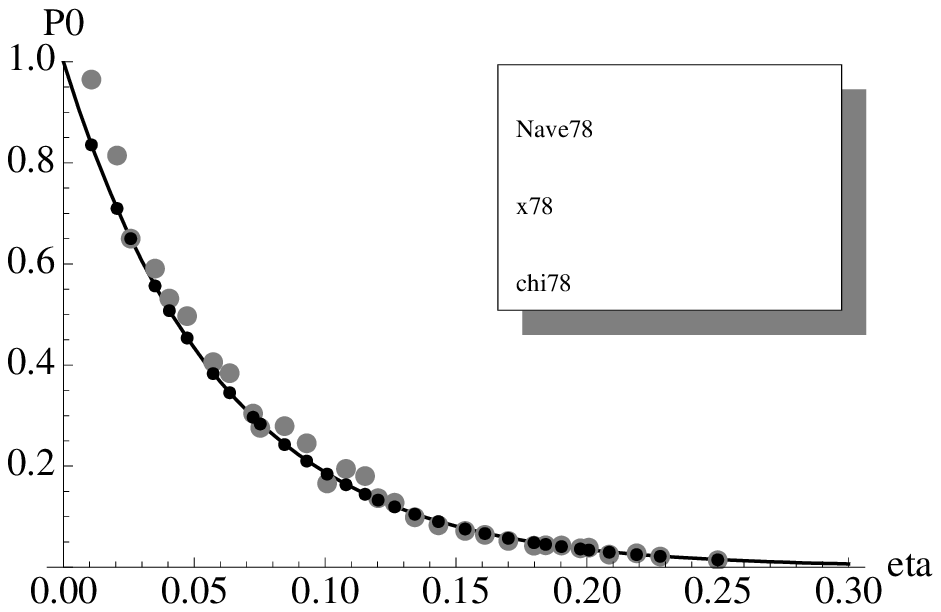}\hspace{0.5cm}
\includegraphics[width=0.45\textwidth]{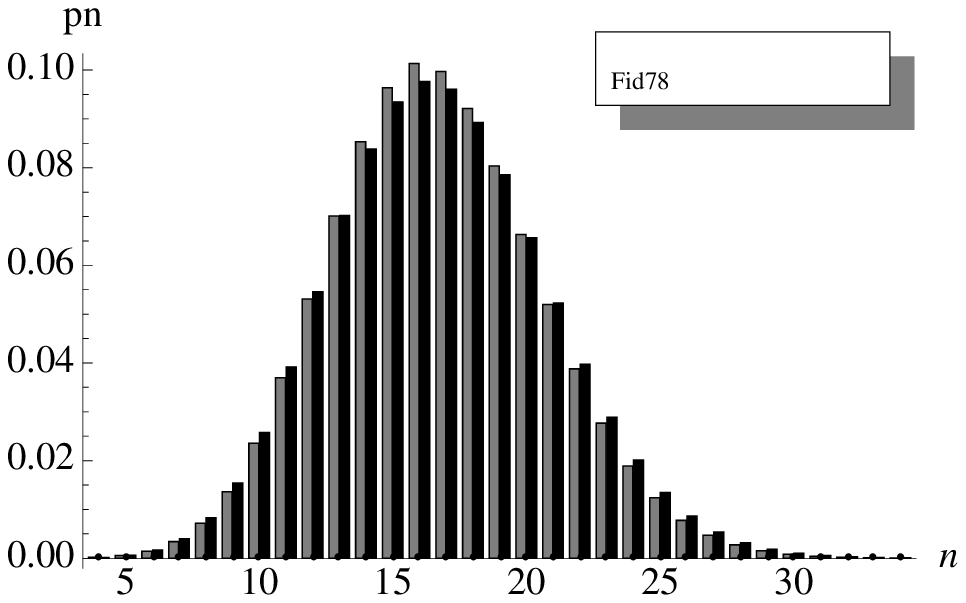}
\vspace{0.3cm}
\includegraphics[width=0.45\textwidth]{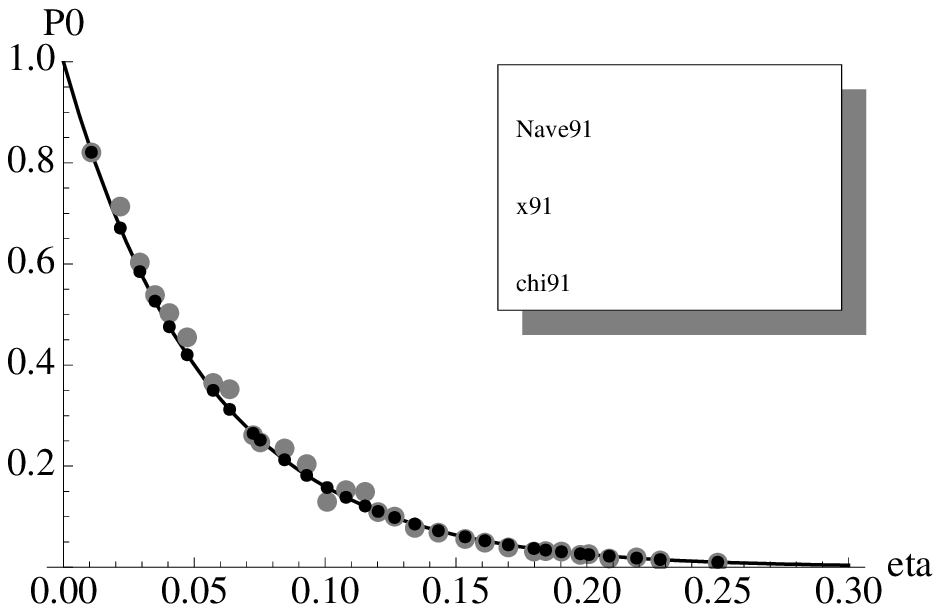}\hspace{0.5cm}
\includegraphics[width=0.45\textwidth]{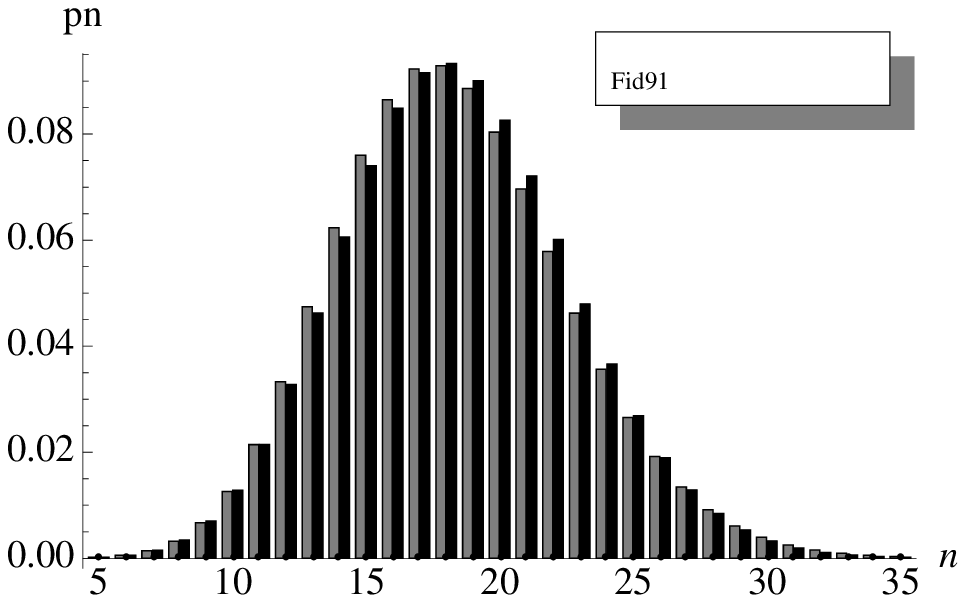}
\end{center}
\caption{On the left: $f_0$ non-click frequencies (gray disks) given by the
stimulated PDC with different stimulation regimes as functions of the
quantum efficiency $\eta$. The black disks are
the ``off'' probabilities obtained by means of the MaxLik reconstructed
photon distribution; the solid line corresponds to
Eq.~(\ref{P0convThD}) with $N_{\rm th}= (1-x) N_{\rm ave}$,
$|\alpha|^2 = x N_{\rm ave}$ and $M=7\,10^{5}$. In each plot we report also the
average number of photons
($N_{\rm ave}$) obtained by the fit of the experimental $f_0$,
the percentage of stimulated emission $x$ and the $\chi^2$ of the
the MaxLik fit.
On the right: MaxLik reconstructed photon distribution (gray bars)
and photon distribution given by Eq.~(\ref{DTpd}) with the same
values of the parameters given in the respective left plots.
In each plot we report also the fidelity $F$ between the two photon
distributions. Note the different ranges of $n$.}
\label{f:results}
\end{figure}
\par
The obtained results are shown
in Figure \ref{f:results}: for the reconstruction we used the
MaxLik estimation with constraint on the energy, as described in the previous
section. The plots on the left show the $f_0$ non-click frequencies given by
the stimulated PDC with different stimulation regimes vs. the
quantum efficiency $\eta$, and the fit obtained by means of the
MaxLik estimation and Eq.~(\ref{P0convThD}). The $\chi^2$ quantity reported
has been defined as the sum of the square differences between
the ``off'' probabilities given by the reconstructed photon statistics
and the measured $f_0$. To quantify the similarity between the two
photon distributions appearing in the plots on the right, instead, we used the
fidelity formula:
\begin{equation}
F=\sum_n\sqrt{\varrho_n^{\rm (ML)}\,\varrho_n^{\rm (DMT)}},
\end{equation}
where $\varrho_n^{\rm (ML)}$ is the MaxLik reconstructed photon distribution
and $\varrho_n^{\rm (DMT)}$ is the one obtained from Eq.~(\ref{DTpd}).
\begin{figure}[htbp]
\psfrag{P0}{\small $f_0$}
\psfrag{eta}{\small $\eta$}
\psfrag{pn}{\small $\varrho_n$}
\psfrag{n}{\small $n$}
\psfrag{NaveNB}{\tiny $N_{\rm ave}=17.55$}
\psfrag{xNB}{\tiny $x=90.7\%$}
\psfrag{chiNB}{\tiny $\chi^2=6.1\, 10^{-3}$}
\psfrag{FidNB}{\tiny $F=92.81\%$}
\begin{center}
\hfill
\includegraphics[width=0.45\textwidth]{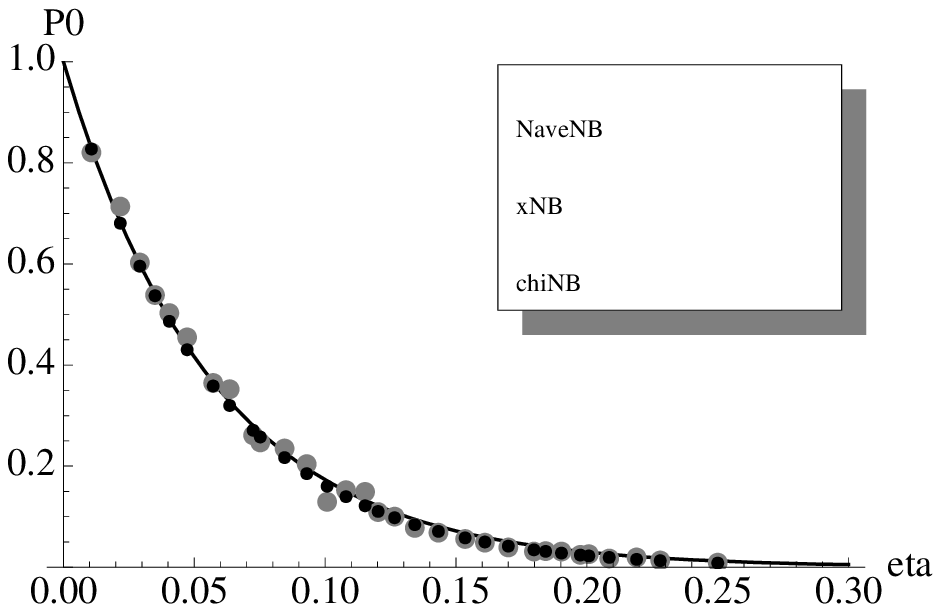}
\hfill
\includegraphics[width=0.45\textwidth]{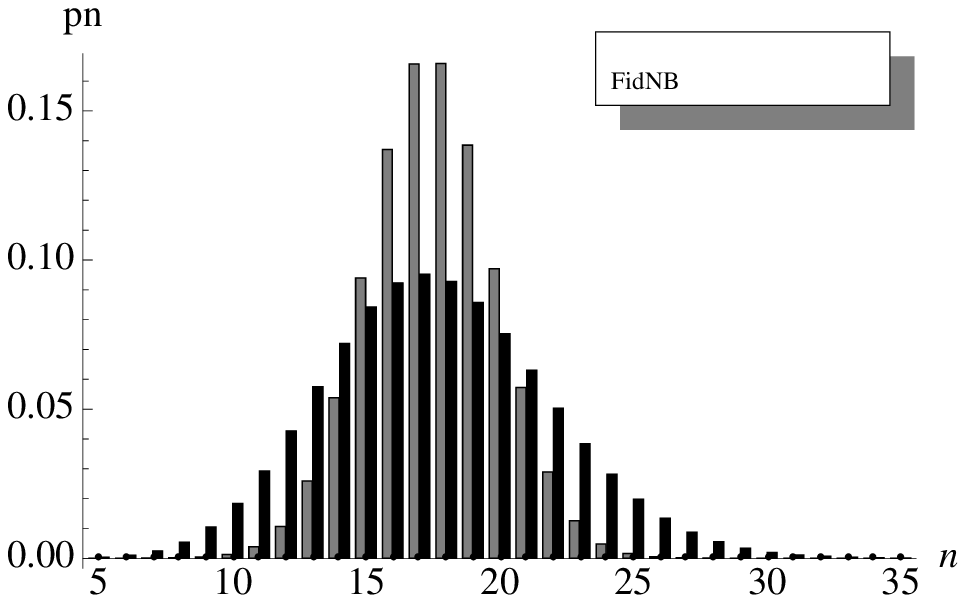}
\hfill
\end{center}
\caption{Same plots as the bottom ones in Figure \ref{f:results}, obtained
without the energy constraint on the algorithm (notice that here $N_{\rm ave}=17.55$
whereas previously it was $N_{\rm ave}=18.25$): even if the MaxLik fit of
the experimental $f_0$ set is good, the fidelity between the MaxLik
reconstructed photon distribution and the one given by
Eq.~(\ref{DTpd}) is quite low.} \label{f:BAD}
\end{figure}
In Figure~\ref{f:BAD} we consider the same scenario giving the bottom
plots of Figure \ref{f:results}, but now we try to perform the
reconstruction without any constraint on the energy ($\beta=0$). We can
see that, even if the MaxLik fit of the $f_0$ frequencies (left side) is
good, the fidelity between the MaxLik reconstructed photon distribution
and the one given by Eq.~(\ref{DTpd}) is quite low (right side): a
result that confirms the advantage of using the constrained MaxLik
method.
\section{Concluding remarks}\label{s:remarks}
In this paper we have shown how an important improvement on the
convergence of the photon statistics reconstruction code, based
on MaxLik estimation applied to on/off detection data, can be
achieved by increasing the number of Lagrange multipliers when some
"a priori" knowledge of the state is available. In particular we
have addressed the reconstruction of the full photon distribution of
multimode fields generated by seeded parametric down-conversion,
demonstrating the advantages of the constrained MaxLik method. This
achievement represents an important step in view of widespread
application of this scheme.
\section*{Acknowledgements}
This work has been partially supported by the CNR-INFM convention,
by Regione Piemonte E14 contract and by 07-02-91581-ASP.

\end{document}